\documentclass[twocolumn,showpacs,preprintnumbers,amsmath,amssymb, pra, aps, 10pt] {revtex4-1}

\usepackage{graphicx}
\usepackage{dcolumn}
\usepackage{bm}
\usepackage{color}
\usepackage{ulem}

\newcommand{\nn}{\nonumber\\}

\newcommand{\ket}[1]{\left|#1\right>}

\newcommand{\bea}{\begin{eqnarray}}
\newcommand{\ea}{\end{eqnarray}}
\newcommand{\eea}{\end{eqnarray}}

\newcommand{\sumint}[1]

\begin{document}



\title{Interaction-induced coherence among polar bosons stored in triple-well potentials}

\author{Bo Xiong and Uwe R. Fischer}
\affiliation{Seoul National University, Department of Physics and Astronomy \\  Center for Theoretical Physics, 
151-747 Seoul, Korea}

\date{\today}

\begin{abstract}

   We study first-order spatial coherence for interacting polar bosons trapped in triple-well potentials.
It is argued that besides the well-known coherence produced by couplings related to tunneling between the sites,  
there exists a nonlocal coherence predominantly determined by intersite interactions, which prevails between the outer sites of the triple well when their total filling is odd. We find that the 
nonlocal interaction-induced coherence originates from the superposition of degenerate many-body states in symmetric triple wells, and demonstrate its robustness against perturbations due to various tunneling mechanisms and thermal fluctuations.
\end{abstract}


\maketitle
      
   \section{Introduction}   
Research into the many-body physics of interacting particles stored in multiple-well geometries has wide-ranging implications for the cross-fertilization of condensed-matter physics and quantum optics  \cite{ahufinger}. 
Preparing ultracold polar molecules in such external scalar potentials possessing local minima,   offers manifold opportunities to investigate the intricate phase diagrams related to the complex coherence properties implied by the long-range dipolar interactions between the molecules. 
In the simplest version of the Hubbard model with contact interactions only, the existence of first-order coherence is conventionally associated with the competition between kinetic (tunneling) energy and two-body interactions. 
By contrast, for polar molecules stored in multi-well geometries, the existence of several energy scales associated with the nonlocal two-body interactions vastly enriches the structure of the phase diagram, and enables the creation of many phases with distinct coherence properties \cite{Trefzger}.

%

   
For the observation of the effects of nonlocal interactions, a triple-well potential is 
the physically most transparent and easily realizable setup to display nonlocal features which go beyond those of contact interactions. A major advantage of the triple well is that it is tractable numerically, e.g., for the ground state by exact diagonalization (ED), when the field operator expansion is restricted to the lowest-lying orbital per site, so that only three field-operator modes are involved in the many-body problem, in distinction to an
optical lattice with a large number of sites. 
In optical lattice terminology, restriction to one orbital per site would be referred to as the single-band Bose-Hubbard model \cite{HubbardToolBox}.
 
Previous theoretical work has explored bosonic triple-well systems 
for contact interactions by a semiclassical approach \cite{moss}, mean-field treatments \cite{grae, buon}, and the multiconfigurational time-dependent Hartree method for bosons \cite{stre}. 
Transistor-like effects in a triple well, where the occupation of the middle well controls tunneling between the outer wells were demonstrated in \cite{Stickney}. 
Polar bosons in triple-well potentials, including however only the conventional single-particle tunneling mechanism, have been investigated by many-body \cite{laha,anna,chatterjee} and mean-field methods \cite{pete}.

   
In what follows, we demonstrate that a triple well containing polar bosons with dipolar interactions  can exhibit variants of the first-order coherence of matter waves of physically distinct origin.
Besides the coherence due to single-particle and occupation-number-weighed tunneling we show that, 
in the appropriate parameter regime, the intersite interaction can generate macroscopic first-order coherence between the two outer sites of the triple well, in the presence of an very small tunneling which conventionally cannot lead to any first-order coherence. 
       We demonstrate both the origin of this interaction-induced coherence and its stability against perturbations due to various tunneling mechanisms (both short-range and long-range) as well as pair exchange. We also point out that a destabilization 
       of the interaction-induced coherence by a possible asymmetry of the trap potential can be counteracted  by a small amount of occupation-number-weighed tunneling, which stems from interactions. We analyze the finite-temperature behavior of the interaction-induced coherence,  
and propose a procedure for its measurement. 
   
   \section{Triple-well Hamiltonian}

      We consider $N$ dipolar bosons trapped in 
      triple-well potentials where both the potential depth in each well and the relative orientation of the polarized dipoles 
is adjustable \cite{laha, pete}. It is furthermore assumed that the lowest-energy orbital 
wavefunction in each well, $\phi_{i}(\mathbf{r})$, where $i$ denotes the well index, 
is fixed and independent of the particle numbers per site (the filling). 
The dipole moment of the bosons is sufficiently strong in the sense
that a large dipole coupling dominates the contact interaction coupling; this is realistically obtained with ultracold polar molecules with large permanent electric dipole moment \cite{bara, laha1}. Expressing the bosonic field operator as $\hat{\psi}(\mathbf{r}) = \sum_{i=1}^{3} \phi_{i}(\mathbf{r}) \hat{a}_{i}$, the low-energy Hamiltonian for the triple well is of the 
single-band Bose-Hubbard form 
   \begin{equation} \label{H}
      \begin{split}
         \hat{H} & =  - J_{1}[\hat{a}_{2}^{\dag}(\hat{a}_{1} + \hat{a}_{3}) + \mathrm{h.c.}] 
                   - J_{2} [\hat{a}_{2}^{\dag}(\hat{n}_{2} + \hat{n}_{1}) \hat{a}_{1} + \mathrm{h.c.}] \\
                 & - J_{2} [\hat{a}_{2}^{\dag}(\hat{n}_{2} + \hat{n}_{3}) \hat{a}_{3} + \mathrm{h.c.}]  
                   - \frac{J_{2}}{\lambda} [\hat{a}_{1}^{\dag}(\hat{n}_{1} + \hat{n}_{3}) \hat{a}_{3} + \mathrm{h.c.}] \\
                 &  + \frac{W}{2} [(\hat{a}_{2}^{\dag})^{2} (\hat{a}_{1}^{2} + \hat{a}_{3}^{2}) + \mathrm{h.c.}]     + \frac{W}{2\lambda} [ (\hat{a}_{1}^{\dag})^{2}\hat{a}_{3}^{2} + \mathrm{h.c.}]   \\
                 & + \frac{U}{2}\sum_{i=1}^{3} \hat{n}_{i} (\hat{n}_{i} - 1) + V \left[ \hat{n}_{1}\hat{n}_{2} + \hat{n}_{2}\hat{n}_{3} + \frac{1}{\lambda} \hat{n}_{1} \hat{n}_{3} \right] \\
                 & +\sum_i \epsilon_i \hat n_i
      \end{split}
   \end{equation}   
Here $J_{1} = -\mathbf{\int} d \mathbf{r} \phi_{i}^{*}(\mathbf{r})\left[-\nabla^{2}/2 + V_{\mathrm{trap}}(\mathbf{r})\right] \phi_{i+1}(\mathbf{r})$ 
(with $V_{\mathrm{trap}}$ the triple well scalar potential and we set $\hbar=m=1$, where 
$m$ is the boson mass), 
$J_{2} = -\mathbf{\int} |\phi_{i}(\mathbf{r})|^{2} \phi_{i}^{*}(\mathbf{r}') \phi_{i+1} (\mathbf{r}') 
V_{\mathrm{dd}}(\mathbf{r} - \mathbf{r}') d \mathbf{r} d \mathbf{r}'$, 
and $W = \mathbf{\int}  \phi_{i}^{*}(\mathbf{r}) \phi_{i}^{*}(\mathbf{r}') \phi_{i+1} (\mathbf{r}) \phi_{i+1} (\mathbf{r}') 
V_{\mathrm{dd}}(\mathbf{r} - \mathbf{r}')  d \mathbf{r} d \mathbf{r}'$  
are the single-particle,  occupation-number-weighed, and pair-tunneling rates respectively \cite{adam, toma}. The coupling $U = g \int |\phi_{i}(\mathbf{r})|^{4} d \mathbf{r} + \int |\phi_{i}(\mathbf{r})|^{2} |\phi_{i}(\mathbf{r}')|^{2} V_{\mathrm{dd}}(\mathbf{r} - \mathbf{r}') d \mathbf{r} d \mathbf{r}'$ characterizes the on-site interaction, 
where $g = 4\pi a_s$ is the familiar short-range interaction constant with $a_s$ the $s$-wave  scattering length, 
and $V = \int |\phi_{i}(\mathbf{r})|^{2} |\phi_{i+1}(\mathbf{r}')|^{2} 
V_{\mathrm{dd}} (\mathbf{r} - \mathbf{r}') d \mathbf{r} d \mathbf{r}'$ is due to 
dipolar-interaction-induced coupling between sites. We neglect the contribution of the contact interaction to the $J_2, V, W$ terms on the basis of the following considerations:  
     (I) the integrals of $J_{2}$, $V$, and $W$ depend strongly on the overlapping area of orbitals. In the overlap area, these integrals for the dipole-dipole interaction (DDI) can increase significantly with reducing the distance $|\mathbf{r} - \mathbf{r}'|$ in contrast with the contact interaction; (II) 
     For heteronuclear molecules with large electric dipole moment, e.g., KRb, ND$_3$ and HCN [14], the dipole coupling length will generally be much larger than the $s$-wave scattering length.
     The ratio of dipolar and contact coupling strengths, $g_d/g$, can then be up to order $10^{2}$ (for the definition of the dipolar coupling $g_d$ see the next paragraph). Finally, $\epsilon_{i}$ is the single-particle energy at site $i$ and we choose $\epsilon_{3} \equiv 1$ as the energy scale.  Alternatively, natural energy units can also be used.
Note that for our purposes the more conventional way of choosing, e.g., the 
single-particle tunneling $J_1$ as unit of energy is not suitable, because we consider below cases where the tunneling exactly vanishes. 

The dipolar interaction between polarized dipoles at positions $\mathbf{r}$, $\mathbf{r}'$ is given by $V_{\mathrm{dd}} (\mathbf{r}-\mathbf{r}') = 3g_d (1 - 3 \mathrm{cos}^{2} \theta )/(4\pi|\mathbf{r} - \mathbf{r}'|^{3})$, where $\theta$ is the 
angle between $\bm r-\bm r'$ and the common direction of the dipoles.  The dipolar coupling constant reads $g_d= \mu_0 d_m^2/3 $ for magnetic and 
$g_d=d_e^2/3\epsilon_0 $ for electric dipoles of strengths $d_m$ and $d_e$, respectively. 
In the units  
which we employ, $g_d$ has dimension of length like the contact coupling $g$.

From point-like, tightly localized orbitals $\phi_{i}(\mathbf{r})$ to two infinite chains of dipoles where the lateral extension of two aligned orbitals is much larger than their separation (e.g., in Fig.\,\ref{Fig7} below,  the condition $\sigma_{y} \gg \sigma_{z}$ amounts to infinite chains of dipoles), the factor $\lambda$ for intersite interactions varies from 8 to 4 \cite{laha}. 
In more detail, the ratio $\lambda$ is defined as 
    \begin{equation}
       \lambda = \frac{\int |\phi_{1}(\mathbf{r})|^{2} |\phi_{2}(\mathbf{r}')|^{2} \mathrm{V}_{dd}(\mathbf{r} - \mathbf{r}') d \mathbf{r} d \mathbf{r}'} {\int |\phi_{1}(\mathbf{r})|^{2} |\phi_{3}(\mathbf{r}')|^{2} \mathrm{V}_{dd}(\mathbf{r} - \mathbf{r}') d \mathbf{r} d \mathbf{r}'}.
    \end{equation}
For the orbitals $\phi_{i}(\mathbf{r})$ localized in all directions, $\lambda = 8$, while $\lambda = 4$ if the width of orbitals in the direction orthogonal to their location is much larger than their separation and orbitals in the other direction are still well localized.
The ratios $\lambda$ of the next-nearest-neighbor and  nearest-neighbor sites vary in general somewhat 
for the contributions $J_{2}$, $W$, and $V$, respectively, depending on the geometry of the triple-well potentials and the overlap of orbital wave functions. For simplicity and because it does not affect our main conclusions below, we take their ratio factors $\lambda$ to be equal; we choose $\lambda = 8$. 

We include the pair tunneling $W$, which 
 has been shown to decide on the question of coherent versus fragmented many-body states in general two-mode models \cite{Bader}. On an optical lattice, it was demonstrated that 
pair-tunneling can change significantly the phase diagram of 
polar bosons on an optical lattice as described by the extended Bose-Hubbard model  \cite{toma}.    

Finally, the fundamental object of interest in what follows, the many-body wavefunction, can generally be expanded in a Fock basis, 
   \begin{equation} \label{eq11}
      |\Psi\rangle = \sum_{n_{1}, n_{2}, n_{3}} f(n_{1}, n_{2}, n_{3})|n_{1}, n_{2}, n_{3} \rangle, 
   \end{equation} 
   where $n_1+n_2+n_3=N$. 
   
   \section{Variants of coherence}
   \subsection{Interaction-induced coherence}      
   
\begin{figure}[htbp]
	          \centering
		        \includegraphics[scale = 0.33]{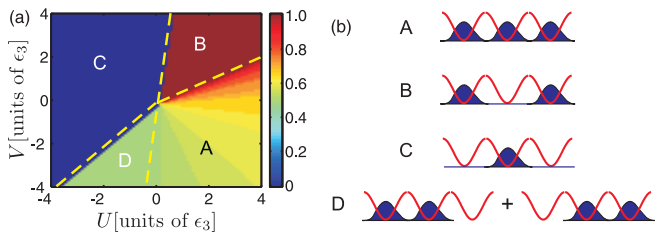} 		        	     
		        \caption{(color online) (a) Phase diagram in terms of the magnitude of $n/N \equiv \langle \hat{n}_1 +\hat n_3 \rangle /N
		        $ as a function of $U$ and $V$ for $N = 30$, $J_{1} = J_{2} = W = 0$, and $\epsilon_{1} = \epsilon_{2} = 1$. The dashed lines in panel (a) show the boundaries between the phases A--D that are shown schematically in (b).} 
	          \label{Fig1}
         \end{figure}

      In the absence of any tunneling, i.e., $J_{1} = J_{2} = W = 0$ and putting $\epsilon_{1} = \epsilon_{2} = 1$,  
      the total energy of the triple-well system subject to the Hamiltonian \eqref{H} reads, 
       \begin{equation} 
      \begin{split}
       \label{MF_eq2}
          E(n_{1},n_{3}) &= (V - U)(N-n_1-n_3) [n_{1} + n_{3}] \\
          &+ \left({V}/{\lambda} - U \right)n_{1} n_{3}.
          \end{split}
       \end{equation} 
We concisely review the phase diagram of polar bosons in a triple well as found in \cite{laha}.
Phase A, cf.\,Fig.\,\ref{Fig1}, is characterized by $n_{1} = n_{3}$, $n_2\neq 0$,  
and appears for $U > 0$ and $V \leq \frac{8}{15} U$, and $U < 0$ and $V < 8U$. Phase B occurs for $n_{1} = n_{3} =  \frac{N}{2}$, $n_2=0$, for $\frac{8}{15}U < V \leq 8 U$ and $U > 0$. The case that $V \geq U$ for $U < 0$, as well as $V > 8U$ for $U > 0$ leads to the phase $C$ where $n_{1} = n_{3} = 0$. Finally, for phase D macroscopic states with $n_{1} = 0$ and $n_{3} \neq 0$ and $n_{1} \neq 0$ and $n_{3} = 0$ are coherently superposed in a Schr\"odinger cat type state, where the regime of parameters is $U < 0$ and $8U < V < U$.

We now point out an important property of phase A, which was not noticed in Ref.\,\cite{laha}. 
We found, as also numerically verified by ED, that when the average fillings in each site, $\bar n_{i}\equiv \langle \hat n_i \rangle$ fulfill 
\bea n \equiv \bar n_{1} + \bar n_{3}\ea 
is odd, so that $\bar n_1$ and $\bar n_3$ are {\it fractional} 
there exist two-fold degenerate states in phase A, with the degenerate subspace spanned by the two states  
\bea
|\Psi_{1}\rangle &=& |(n-1)/2, N-n, (n+1)/2\rangle, \nn
|\Psi_{2}\rangle &=& |(n+1)/2, N-n, (n-1)/2\rangle.\label{psi1psi2}
\ea 
On the other hand, when $n$ is even, 
the many-body state is a single Fock state, $|\Psi\rangle = |n/2, N-n, n/2\rangle$. 
All off-diagonal first-order coherence matrix elements, characterized by $g_{ij}^{(1)} = \frac{1}{2\bar{n}} \left[ \langle \Psi| \hat{a}_{i}^{\dag} \hat{a}_{j}| \Psi \rangle + \mathrm{h.c.} \right]$, are then zero: $g_{12}^{(1)} = g_{23}^{(1)} = g_{13}^{(1)} = 0$. This is, however, not the case for a superposition of the states in Eq.\,\eqref{psi1psi2},
as we will now argue.

\begin{figure}[htbp]
	          \centering
		        \includegraphics[scale = 0.35]{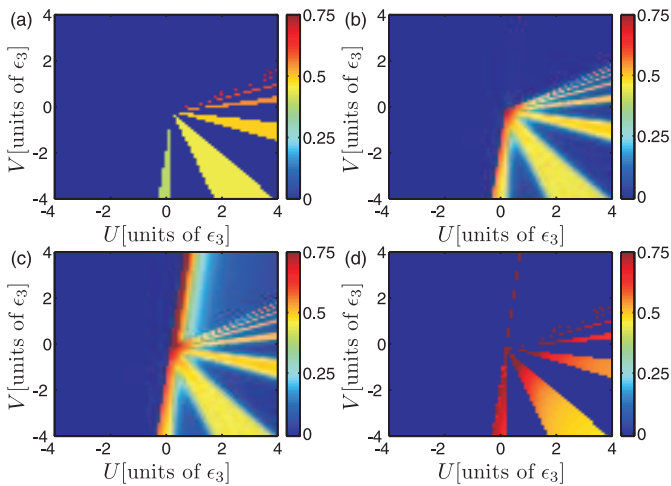} 		        	     
		        \caption{(color online) Illustration of the insensitivity of the ``fan'' structure of the coherence function 
		        $g_{13}^{(1)}$, related to the phase transitions between even and odd values of $n$ 
		        for varying tunneling couplings.  
Magnitude	of $g_{13}^{(1)}$ for (a) extremely small tunneling, $J_1,J_2, W \approx 0$, e.g. 
$J_1=W=0$, $J_2=0.00001$,   
(b) $J_{1} = 0.02$, and $ J_{2} = W = 0$, 
		        (c) $J_{2} = 0.001$, and $J_1=W = 0$, (d) $W=-0.02$, and $J_1=J_2=0$; $N=30$ and $\epsilon_{1} = \epsilon_{2} = 1$ in all plots.}  
		        \label{Fig2}
	          \end{figure}

The many-body state is in general any linear combination of the two degenerate states, i.e., $|\Psi\rangle = \alpha |\Psi_{1}\rangle + \beta |\Psi_{2}\rangle$ where the coefficients $\alpha$ and $\beta$ can be any complex number and fulfill 
$|\alpha|^{2} + |\beta|^{2} = 1$. However, for {\it small } tunneling in a realistic system, e.g., any small $J_{1}$, $J_{2}$, or $W$, 
$\alpha$ and $\beta$ adjust to relative phase zero. Moreover, the spatial symmetry between sites 1 and 3 requires $\bar n_{1} = \bar n_{3}$. 
The only possible form of the many-body wavefunction then is 
        $  |\Psi\rangle = \frac{e^{i\theta}}{\sqrt{2}} [|\Psi_{1}\rangle + \Psi_{2}\rangle],$
with some global phase $\theta$. Setting the latter to zero, we conclude that 
   \begin{equation}
         |\Psi\rangle = \frac{1}{\sqrt{2}}\left[ |\Psi_{1}\rangle + |\Psi_{2}\rangle \right]. \label{superpos} 
   \end{equation} 
The above many-body state induces macroscopic  nonlocal first-order coherence, i.e., the quantity   
\bea 
g_{13}^{(1)} = \frac{1}{2\bar{n}} \left[\langle \Psi| \hat{a}_{1}^{\dag} \hat{a}_{3} |\Psi\rangle + \mathrm{h.c.} \right]
= \frac{n + 1}{4\bar n} \label{g13}
\ea 
is of order unity, cf.\,Fig.\,\ref{Fig2}, where $\bar n = \sum_{i=1}^3 \bar n_i/3$. 

We emphasize that (infinitesimally) small tunneling couplings, which conventionally can not lead to any finite coherence in the system, are a necessary (but relatively minor) condition for the nonlocal coherence induced by interactions to be established. We propose that the optimal choice of tunneling here, to realize the interaction-induced coherence, is to employ a certain amount of occupation-number-weighed tunneling. For single-particle tunneling $J_{1}$, a too large hopping amplitude washes out the interaction-induced coherence (also see below), while for a too small $J_1$ thermal fluctuations would destroy the coherence even at very small temperatures because the gap between ground state and the first excited state is order of $J_{1}^{2}$. By contrast, for the number-occupation-weighted tunneling, $J_{2}$, the gap between the ground state, including the nonlocal interaction-induced coherence, and the lowest excited state without the coherence, is of the order of $\frac{J_{2}}{\lambda}  n^{2}$. This implies that the interaction-induced coherence is more realistically obtained by having finite $J_{2}$, since the corresponding gap depends strongly on the square of $n$. To keep $n$ large and avoid the potential washing out of the coherence fan structure (see also below), it is preferable to keep the occupation number in the middle site much less than on the outer sites, for example, by making the single particle energy in the middle site larger than on the outer sites, $\epsilon_{2} > \epsilon_{1} = \epsilon_{3}$. In this situation, the coherence between the nearest-neighbor sites caused by $J_{2}$ is negligibly small.     

We, furthermore, stress that the interaction-induced coherence only exists between the outer sites of the triple well, i.e., $g_{12}^{(1)} = g_{23}^{(1)} = 0$ and, according to Eq.\,\eqref{g13}, $g_{13}^{(1)} = \frac{n+1}{4\bar{n}}$. The single-particle density matrix, $\langle \hat{a}_{i}^{\dag} \hat{a}_{j} \rangle$, indeed has three macroscopic eigenvalues, which are 
which are $\bar n_2$ and $(\bar n_{1} + \bar n_{3})/2 \pm \sqrt{5 \bar n_{1}^{2} + 5 \bar n_{3}^{2} - 6 \bar n_{1} \bar n_{3} 
+ 2 \bar n_{1} + 2 \bar n_{3} + 1}/{4}$.
When $n\gg 1$ (but $n\le N$), the three eigenvalues are approximately $\frac{3n +1}{4}$, $\frac{n-1}{4}$, and $N - n$. 
Note that for the proposed parameter regime, $U > 0$ and $V < 0$ in phase A, $\frac{N}{2} < n < \frac{2N}{3}$; hence there exist at least two macroscopic eigenvalues with a minimum value larger than $\frac{N}{8}$. This demonstrates that, seen from the point of view of the {\it entire} triple well, the superposed states with macroscopic interaction-induced coherence between the outer sites, are globally speaking fragmented condensate states \cite{penr}.

Finally, we point out that the superposition \eqref{superpos} of two-fold degenerate states is much more robust than the degenerate states occurring in phase C.
For vanishing tunneling, the many-body wavefunction in the parameter regime of phase C should be superposed from three degenerate states, i.e., $|\Psi\rangle = a |N, 0, 0 \rangle + b|0, N, 0 \rangle + c|0, 0, N\rangle$ where any two of $a$, $b$, and $c$ are zero and the remaining one unity. But this trifold degeneracy is broken readily against a small perturbation of first-order coupling, e.g., the single-particle tunneling between wells, and the many-body wavefunction favors the single-particle state $|0, N, 0\rangle$. By contrast, a small perturbation due to tunneling
does not destroy the degeneracy of $|\Psi_{1}\rangle$ and $|\Psi_{2}\rangle$, as we will demonstrate now. 

The emergence of the first-order coherence for $n = \mathrm{odd}$ and disappearance for $n = \mathrm{even}$ 
results in the formation of a fan structure of the first-order coherence in the parameter regime of phase A, see Fig.\,\ref{Fig2}. 
The fan leaves with nonlocal interaction-induced coherence residing in them indicate that the superpositions vary in a regular manner \cite{maik} 
with respect to the parameters, $U$ and $V$ and with $n = \mathrm{odd}$. 
As an example, fixing $V$ and increasing $U$, one obtains, for $n = 17$, the superposition $\frac1{\sqrt{2}}(|9, 13, 8 \rangle + |8, 13, 9 \rangle)$ and subsequently for $n = 15$, $\frac1{\sqrt{2}}(|8, 15, 7 \rangle + |7, 15, 8\rangle)$, and so on.

To explain the structure of the phase diagram in region A of the phase diagram in more detail, 
we now argue that while small single-particle, occupation-number-weighed, 
or pair tunneling couplings slightly broadens the statistical distribution of the many-body wavefunction \eqref{eq11} 
around the superposed state in Fock space, they do not change the structure of the phase diagram. For concreteness, with a finite but small $J_{1}$, the many-body wavefunction should be of the form, 
   \begin{equation} \label{Coh_eq1}
      \begin{split}
         |\Psi\rangle & = c_{1} | (n-1)/2, N-n, (n+1)/2 \rangle \\
                      & + c_{2} | (n+1)/2, N-n, (n-1)/2 \rangle \\
                      & + c_{3} | (n-3)/2, N-n, (n+3)/2 \rangle \\
                      & + c_{4} | (n+3)/2, N-n, (n-3)/2 \rangle \\
                      & + \mathrm{other ~ terms}\ldots,
      \end{split}
   \end{equation}     
where the additional single-particle states are determined by $J_{1}$ and the larger $J_{1}$, the more single-particle states need to be included. For small $J_{1}$, $c_{1}$ and $c_{2}$ are dominant, and determined by $U$ and $V$, such that the interaction-induced coherence is obtained. Only with strongly increasing $J_{1}$, the 
contributions of more single-particle states broaden the distribution of the many-body state in Fock space,  
significantly reducing the value of $c_{1}$ and $c_{2}$, and then suppressing the interaction-induced coherence. 
A similar discussion can be applied to the couplings $J_{2}$ and $W$. Therefore, effects caused by the various tunnelings do not result in an essential change for the  interaction-induced coherence, as long as the tunneling
perturbations remain sufficiently small. 
This robustness of the coherence against perturbations has been numerically verified by the the ED calculations shown in Fig.\,\ref{Fig2}. It can be concluded from Fig.\,\ref{Fig2} that the magnitude of the coherence  changes due to finite tunnelings of various origin. However, the phase boundaries (the ``fan'' structure), while 
slightly blurred by small tunneling, remain basically  intact. Note for the proper interpretation of the parameter values chosen that the relevant quantities to compare with each other are $J_1$ and $nJ_2$. 

By comparing Figs.\,\ref{Fig2}\,(b) and (c) we see that the nature of the crossover between the phases B and C is strongly influenced by  number-weighed tunneling, while ordinary tunneling has basically no effect on the transition. This highlights the long-range nature of the tunneling mechanism correlated to site occupation, embodied in the $J_2$ terms. In addition, it furnishes a possible method to {\it measure} $J_2$ independently from $J_1$ in the triple well by investigating the transition region between phases B and C. 

We finally conclude from Fig.\,\ref{Fig2}\,(d) that the effect of pair-exchanges $\propto W$ on the 
{\it single-particle} coherence magnitude is most pronounced, confirming the profound effect pair exchanges have on the first-order coherence \cite{Bader}.


  
\subsection{Small bias energies} 

When a small bias energy $\delta E = \epsilon_{1} - \epsilon_3$ between the outer sites  
(chosen here to be negative) is accounted for, the actual triple-well ground state should be $\ket{\frac{n+1}{2}, N-n, \frac{n-1}{2}}$ and the associated energy is 
$(V-U)(N-n)n + \left( \frac{V}{\lambda} - U \right) \frac{n^{2} -1}{4} + \delta E (n+\frac{1}{2})$ while the energy of the superposition state with respect to the Hamiltonian \eqref{H}) is $(V-U)(N-n)n + \left( \frac{V}{\lambda} - U \right) \frac{n^{2} -1}{4} + \delta E n$ (the superposition \eqref{superpos} 
is not the eigenstate of the Hamiltonian \eqref{H}). 
Therefore the superposition  energy is higher than the ground-state energy by $|\delta E|/2$, indicating that for any asymmetry between sites 1 and 3, the two-fold degeneracy and therefore the coherence is broken. However, for a  negative intersite interaction, i.e., $V < 0$, choosing Gaussian orbitals (also see section \ref{parameters} 
below)
shows that $J_{2}$ is positive. This implies that $J_{2}$ 
stabilizes the interaction-induced coherence states. For a \textit{small} $\delta E < 0$ and $J_{2} > 0$, the ansatz for the many-body state can be written 
$\eta \ket{\frac{n+1}{2}, N-n, \frac{n-1}{2}} + \gamma \ket{\frac{n-1}{2}, N-n, \frac{n+1}{2}}$,  
where $\eta$ and $\gamma$ are real and have equal sign, with normalization $\eta^{2} + \gamma^{2} = 1$.
The energy contributed by the bias $\delta E$ is $\left[ \frac{n}{2} + \frac{1}{2} (\eta^{2} - \gamma^{2}) \right] \delta E$, while that of $J_{2}$ is $-\frac{J_{2}}{\lambda} \eta \gamma n(n+1)$. It is then straightforward to obtain that the condition for stabilizing the interaction-induced coherence 
state is given by $J_{2} > \frac{\lambda |\delta E|}{n + 1}$ for not too large $|\delta E|$ (for increasing $J_2$, more single-particle states
need to be included).  This condition has been numerically confirmed as well.
  \subsection{Detection of interaction-induced coherence}

  \begin{figure}[t]
	          \centering
		        \includegraphics[scale = 0.5]{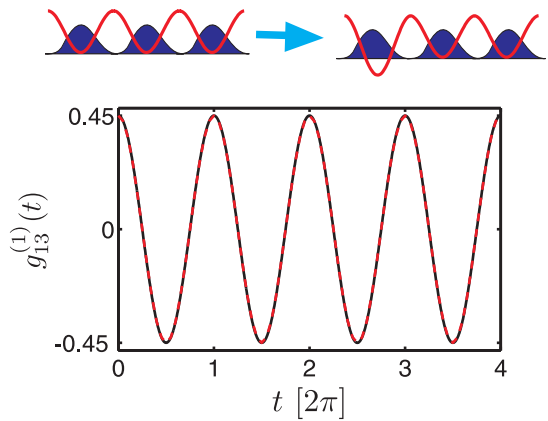} 		        	     
		        \caption{(color online) Time evolution of $g_{13}^{(1)}(t)$ for an initial state with $U = 1.5$, $V = -2$, $J_{2} = 0.00001$, $J_{1} = W = 0$ and $\epsilon_{1} = \epsilon_{2}$, quenched into $\epsilon_{1} = 0$ and identical other parameters (see schematic representation on top).
		        The black solid line is from a 
		        numerical solution to the many-body equation of motion, and the red dashed line from formula (\ref{MF_eq5}).} 
	          \label{Fig3}
         \end{figure}        
      

We propose to verify the existence of interaction-based coherence with the nonequilibrium dynamics of the quantum many-body state. The initially assumed superposition state is assumed to include the coherence between the outer sites but no coherence between the nearest-neighbor sites. It is quenched towards a nonequilibrium state by breaking suddenly the symmetry of sites 1 and 3, 
applying a single-energy site bias to site 1 instantaneously, cf. the illustration in Fig.\,\ref{Fig3}. 
The time-dependent many-body wave function
is then of the form, for sufficiently small $J_2$, 
    \begin{widetext}  
       \begin{equation} \label{MF_eq7}
             |\Psi (t) \rangle  \simeq 
             \frac{1}{\sqrt{2}} \mathrm{exp}\left[-it\left((V - U)(Nn - n^2)  
            +\left(\frac{V}{\lambda} - U\right) \frac{n^{2} - 1}{2} + N \right) \right]
            \!\sum_{\gamma=\pm}\exp\left[-it \delta E \frac{n+\gamma}2 \right]\ket{\frac{n + \gamma}{2},N-n, \frac{n - \gamma}{2}} 
       \end{equation}        
    \end{widetext}
    where $n$ is assumed to be odd.    
Therefore, it is obtained readily that  
      \begin{align} 
            g_{13}^{(1)} & = \frac{1}{2\bar{n}} \langle \Psi(t)| (\hat{a}_{1}^{\dag} \hat{a}_{3} + \mathrm{h.c.}) |\Psi(t) \rangle \nonumber \\
                         & = \frac{(n+1)}{4 \bar{n}} \mathrm{cos} [\delta E t]  \label{MF_eq5} \\
            g_{12}^{(1)} & = g_{23}^{(1)} = 0 \label{MF_eq6}.
      \end{align}            
Formulas (\ref{MF_eq5}) and (\ref{MF_eq6}) show that there exists {exclusively} the time-dependent phase coherence
between sites 1 and 3, for which the out-of-equilibrium 
oscillation frequency is determined by $\delta E$, in complete agreement with our numerical results displayed in Fig.\ref{Fig3}.     
The interaction-induced coherence therefore can be measured by means of atomic interference, where at different instants of evolution with respect to the same initial setup, the trap potentials are released such that the freely expanding clouds from sites 1 and 3 will interfere. The visibility of the interference pattern then amounts to a measure of the first-order coherence \cite{gerbier}. The quasi-momentum distribution function (structure
factor) for the freely expanding clouds reads
   \begin{equation} \label{MF_eq77}
      n_{k}(t) = \frac{1}{3} \sum_{i, j} \langle \hat{a}_{i}^{\dag} \hat{a}_{j} \rangle e^{i k l (i - j)},
   \end{equation}  
where $l$ is the separation of the nearest-neighbor sites. For the initial superposition state, $n_{k} = \frac{N}{3} + \frac{n+1}{6} \mathrm{cos}(2 k l)$. On the other hand, for the triple-well coherent state,  $n_{k} = \frac{N}{3} \left[1 + \frac{4}{3} \mathrm{cos}(k l) + \frac{2}{3} \mathrm{cos}(2 k l) \right]$. Therefore, when the occupation number in the outer sites is sufficiently large, the contrast of interference from the superposition should be manifest, when comparing with the fully coherent state. In the extreme case of $n = N$ where the atomic occupation in site 2 is completely suppressed, the structure factor shows the largest possible contrast of interference coming from the nonlocal interaction-induced coherence, which is approximately $33\%$ of that of a fully coherent state. This can be measured by time-of-flight experiments.

   \subsection{Coherence versus temperature}    

\begin{figure}[t]
	          \centering
 		        \includegraphics[scale = 0.39]{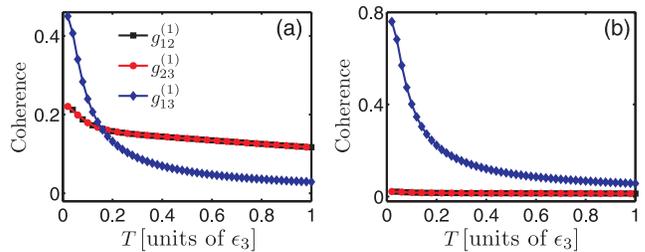} 		        	     
		        \caption{(color online) The first-order spatial coherence, $g_{ij}^{(1)}(T)$ as a function of temperature. (a) All $\epsilon_{i} = 1$ for $i = 1, 2, 3$, $U = 2.5$, $V=-3$, $J_{2} = 0.001$, $N = 30$. (b) strongly reduced occupation of the central site, using  $\epsilon_{1} = 1$, $\epsilon_{2} = 114.75$, and other parameters identical to (a).} 
	          \label{Fig6}
         \end{figure}

      Here, we explore the effect of finite temperatures on the interaction-induced coherence. 
      In the canonical ensemble, the thermal average of an operator $\hat{O}$ is 
      $\langle \hat{O} \rangle = \sum_{j} \frac{e^{- E_{j}/T}}{Z} \langle j |\hat{O}| j \rangle$,  
   where the canonical partition sum $Z = \sum_{j} e^{-E_{j}/T}$ and $| j \rangle$ are the energy eigenstates. Fig.\,\ref{Fig6} shows that in the regime of small temperatures, 
   the interaction-induced coherence, i.e.,  $g_{13}^{(1)}(T)$, is much larger than the coherence caused by tunneling couplings. It should be noticed that in Fig.\ref{Fig6}(a) the mean occupation numbers in each site are approximately equal, $\bar{n}_{1} = \bar{n}_{3} = 8.5$, and $\bar{n}_{2} = 13$, and thus the interaction-induced coherence can be significantly enhanced and the conventional coherence can be suppressed greatly if the mean occupation in the central site is considerably reduced, e.g., by adjusting the single-particle energy in center well to be much larger than in the outer wells, $\epsilon_{2} \gg \epsilon_{1} = \epsilon_{3}$. 
      Correspondingly, the transition temperature below which the interaction-induced coherence dominates increases significantly. We illustrate this fact in Fig.\,\ref{Fig6}\,(b) where,
due to large $\epsilon_{2}$, the occupation number in the central site 2 is suppressed to be 
close to unity,  and the transition temperature strongly increases, approaching the limit for which all  coherence disappears.  
      
      Next we compare 
      the transition temperature with 
energy gaps based on the three lowest energy states in Fig.\,\ref{Fig6}\,(a): the lowest energy state is closely the superposition, i.e., $\frac{1}{\sqrt{2}} [|(n-1)/2, N-n, (n+1)/2 \rangle + |(n+1)/2, N-n, (n-1)/2 \rangle ]$ surrounded by some slightly perturbing states, while the lowest excited state essentially superposes the same two Fock states which are dominating the ground state but with a negative sign, i.e.,  $\frac{1}{\sqrt{2}} [|(n-1)/2, N-n, (n+1)/2 \rangle - |(n+1)/2, N-n, (n-1)/2 \rangle ]$; the third lowest
state is approximately the single Fock state $|(n+1)/2, N-n-1, (n+1)/2 \rangle$, again with the participation of some slightly perturbing states. 
      It can seen readily that the lowest energy gap is mainly contributed by the occupation-number-weighted tunneling 
      and is of order $\frac{J_2}{\lambda} n^2$; the value of this gap corresponding to Fig. 4(a) is 0.038.
      The transition temperature is approximately five times as large as this gap. 
      By contrast, the second lowest energy gap arises from intersite interaction and can be approximated by $(V - U) (N - 2n -1) + \left(\frac{V}{\lambda} -U \right) \frac{n + 1}{2} - \frac{J_{2}}{2\lambda} n (n + 1)$; the corresponding value in Fig.\,\ref{Fig6}(a) is $1.6$, which is about nine times as large as the transition temperature. A similar situation,  namely that the transition temperature is larger than the lowest energy gap but smaller than the second lowest energy gap is also realized in Fig.\,\ref{Fig6}\,(b), where the lowest gap is $0.1$ and the second gap is approximately $2.57$.  This indicates that the nonlocal interaction-induced coherence is robust against thermal fluctuations. Specifically, it persists for temperatures even larger than the lowest energy gap, which is due to occupation-number-weighted tunneling. 
With increasing temperature, the interaction-induced coherence is suppressed gradually. 

Summarizing this subsection, the interaction-induced coherence increases at low temperature significantly, and over the nearest-neighbor coherence,  
further emphasizing the quantum origin of the coherence between the outer sites.            

   \section{Experimental setup and parameter correlations}
   \label{parameters}
        \begin{figure}[b]
	        \centering
		      \includegraphics[scale = 1]{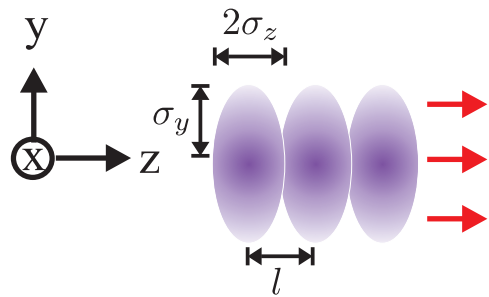} 		        	     
		      \caption{(color online) The configuration of a triple pancake system where 
		      the three 2D pancakes are aligned along the $\hat y$ direction, and all dipoles are oriented along $\hat z$ (red arrows), such that $U>0$ and $V<0$.} 
	         \label{Fig7}
         \end{figure}

We now discuss the experimental feasibility for obtaining the nonlocal interaction-induced  coherence and the associated realistic parameter values for a dipolar gas. 
To achieve the dynamical manipulation of the triple-well bias energy, specifically designed optical potentials can be used. 
One employs a light sheet to provide tight confinement perpendicular to the axis of the focused optical tweezer beam where a single well may be created \cite{laha}. Subsequently, using an acousto-optic modulator to toggle the single well between several positions at high rate to create almost arbitrary time-averaged potentials \cite{hend}, one can adjust the single-energy offsets in time. This setup of the triple-well potentials provides a practical means to control the atom number in each well, which is of critical importance in the realization of interaction-induced nonlocal coherence, e.g., by performing evaporative cooling with different energy offsets in each site. 
Furthermore, one changes the orientation of the external magnetic field to make the DDI direction parallel to a major axis of the 
aligned three condensates, as shown in Fig.\ref{Fig7}, so that  repulsive on-site interaction, attractive intersite interaction, and positive occupation-weighted tunneling are accessible within such a setup. 

While any shape or profile of the condensates is in principle possible for exploring  interaction-induced coherence, a 2D configuration of condensates, cf. Fig. 5, appears most suitable. In contrast to 1D condensates, where strong fluctuations potentially disturb the formation of nonlocal interaction-induced coherence, 2D condensates are much less prone to fluctuations. More importantly, the ratios $-J_{2}/V$ and $W/V$ in the 2D configuration are less susceptible to the width of the condensate in the direction perpendicular to the orientation of DDI (as shown below), which allows to adjust the interaction parameter more easily than in 1D condensates. On the other hand, for 3D condensates, the precise control of the relation between the interaction parameters is more intricate than for a 2D condensate. 
Once the triple-well system is established, one adjusts the height of the potential barriers to suppress the single-particle hopping rate $J_{1}$, which can be reduced exponentially with increasing barrier height.       

It should be observed that it is difficult to detect the fan structure of the interaction-induced coherence if the total particle number $N$ in the triple well is not fixed. The fan structure consists of approximately $N/2$ bright and dark regions (the precise
number depending on the interaction parameters); hence when the particle number changes by just one unit the fan structure and leaf size in Fig.\,\ref{Fig2} changes considerably. Therefore, to employ relatively small $N$ facilitates observation of the fan structure
related to nonlocal interaction-induced coherence.
Single-atom sensitivity in time-of-flight has been achieved, for example with fluorescence imaging, so that even relatively small values of $N$ should in principle be detectable \cite{buecker}.
      A further possibility to increase the signal is to construct a sequence of triple wells in a trichromatic optical lattice setup, which  would, however, incur the difficulty to ensure equal atom numbers for each triple well to ensure that the fan structures in each well
are sufficiently close to each other, differing by less than a particle per triple well. 
This is however not an impossible task for relatively small filling of a few particles in each triple well.

\begin{figure}[t]
	        \centering
		      \includegraphics[scale = 0.55]{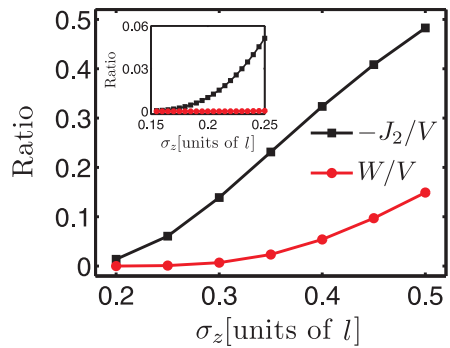} 		        	     
		      \caption{(color online) The ratios $-J_{2}/V$ and $W/V$ for various values of the Gaussian width  
		      $\sigma_{z}$ in units of the well distance, cf.\,Fig.\ref{Fig7}. Filled black squares ($-J_2/V$) and red circles ($W/V$) are numerically evaluated data points, and lines are a guide to the eye.
		      The inset displays a magnified view with higher resolution, of the ratios for highly localized Gaussians (small $\sigma_z$).}
	         \label{Fig8}
         \end{figure}

The triple pancake trap configuration in Fig. 5 achieves $U>0$ and $V<0$.
To understand the parameter relation between $V$, $J_{2}$, and $W$, we  take as the low-energy approximations to the exact Wannier states  orbital wave functions corresponding to the harmonic-oscillator ground-states at each well \cite{mazz}. Specifically the single-particle orbitals are assumed to be of the form, 
         \begin{align}
            \phi_{1} (\mathbf{r}) & = \frac1{\sqrt{2\pi \sigma_{z}\sigma_{y}}}  \mathrm{exp}\left[-\frac{z^{2}}{4\sigma_{z}^{2}} -\frac{y^{2}}{4\sigma_{y}^{2}} \right] \delta(x) , 
            \nn
            \phi_{2} (\mathbf{r}) & =  \frac1{\sqrt{2\pi \sigma_{z}\sigma_{y}}} \mathrm{exp}\left[-\frac{(z-l)^{2}}{4\sigma_{z}^{2}} -\frac{y^{2}}{4\sigma_{y}^{2}} \right] \delta(x) , 
            \label{Ham_eq26} 
            \\
            \phi_{3} (\mathbf{r}) & = \frac1{\sqrt{2\pi \sigma_{z}\sigma_{y}}}  \mathrm{exp}\left[-\frac{(z-2l)^{2}}{4\sigma_{z}^{2}} -\frac{y^{2}}{4\sigma_{y}^{2}} \right] \delta(x). \nonumber
         \end{align}
We take the separation of the two nearest-neighbor orbital wave functions, $l$, as the unit of length and vary $\sigma_{z}$ to obtain the relation of $-J_2/V$ and $W/V$ as shown in Fig.\,\ref{Fig8} (the ratios are independent of $\sigma_y$). 
The numerical results upon evaluating the parameter ratios 
show that $J_{2}>0$ and $W<0$ are consistently smaller than $|V|$, and with increasing overlap of two orbital wave functions, the ratios $-J_{2}/V$ and $W/V$ increase monotonically (see  Fig.\,\ref{Fig8}). 
By contrast, for contact interaction, $W/V$ is constant 
and $-J_{2}/V \gg 1$ for small overlap, {\it decreasing} with increasing overlap \cite{mazz}. This indicates that the dipole-dipole interaction tends to enhance the interaction-induced number-weighed tunneling over the nearest-neighbor interaction coupling for increasing overlap (in the setup of
Fig.\,\ref{Fig7}), while contact interactions have the opposite effect. 

To observe interaction-induced coherence, we assume for concreteness that RbCs molecules with a dipole moment of magnitude $d_e=1.25$ D [24] are trapped in the triple well.
For a numerical example, suppose that a triple-well potential is formed by three Gaussian beams of waist $1\, \mu$m separated by $l = 1.8\, \mu$m, as proposed in Ref.\cite{laha}. For a barrier height $2\pi\times 4000\,$Hz, the single-particle tunneling is 
$J_{1}\sim 2\pi\times 13$\,Hz, so that $N |V|/J_{1}\sim 1.6\times 10^2$ and $N J_{2}/J_{1} \sim 12$ for $N = 30$.

   \section{Conclusion}
   
      We have shown that intersite interaction in a triple well, due to large dipole-dipole interactions, in a broad parameter regime, can generate macroscopic first-order coherence between the two outer sites. This takes place in the presence of very small tunneling, e.g., occupation-weighted tunneling between the three sites. The interaction-induced coherence originates, at zero temperature, from the superposition  of two degenerate states of polar bosons in a symmetric triple-well potential. While globally 
the many-body state is fragmented, it exhibits this partial coherence between the modes localized on the 
outer sites,  
which is robust against tunneling perturbations as well as pair exchange. We have also demonstrated that the fragility of the coherence against site biases (a mismatch 
of energy minima lifting the degeneracy) can be counterbalanced by  occupation-number-weighed tunneling. 
Finally, the interaction-induced coherence is robust against finite temperatures even 
larger than the smallest gap generated by number-weighed tunneling, and up to 
temperatures corresponding to the second gap. This is due to two coherently superposed 
states, separated by the smallest gap, contributing  
to the interaction-induced coherence at finite temperature, as opposed to only one superposed state at zero temperature. The interaction-induced coherence is destroyed by temperatures so large that a third, non-superposed, excited state comes into play.
 
The interaction-induced coherence between sites 1 and 3 executes Josephson oscillations when the initial superposition state is quenched towards a nonequilibrium state by suddenly breaking the symmetry of the triple well (rapidly turning on a site bias), where the associated oscillation frequency is determined by the bias. 
This provides an experimental method to measure the strength of the interaction-induced mesoscopic coherence between the outer sites of the triple well. 

To conclude, we note that a recent work has shown that repulsive interactions induce first-order correlation between spatially separated chains of one-dimensional dipolar fermions \cite{chi}. 
Our origin of coherence is determined by both the spatial symmetry of the triple well and the intersite interaction.
On the other hand, due to the existence of many degenerate states in a very narrow window of intertube interactions in the fermionic system, an infinitely  small intertube tunneling can mix many degenerate states,
 and then lead to a coherent ground state.     
Although nonlocal interactions $V$ 
are the crucial factor in both cases, the interaction in the fermionic case is repulsive, i.e., $V > 0$, in distinction to the present case of elementary bosons, where the first-order coherence arises from negative $V$.
In addition, the 
coherent-state regime for $V$ is very narrow in the fermionic system, 
in contrast to the bosonic triple well, where $V$ and $U$ cover a wide area 
in parameter space. 

\acknowledgments
This research was supported by the NRF Korea, Grants No. 2010-0013103 and No. 2011-0029541, and the 
Seoul National University Foundation Research Expense. 


\begin{thebibliography}{000}
   
   \bibitem{ahufinger} M. Lewenstein, A. Sanpera, V. Ahufinger, B. Damski, A. Sen(De), U. Sen, Adv. Phys. {\bf 56}, 243 (2007).
   
   \bibitem{Trefzger} C. Trefzger, C. Menotti, B. Capogrosso-Sansone, and
M. Lewenstein, 
J. Phys. B: At. Mol. Opt. Phys. {\bf 44}, 193001 (2011). 

\bibitem{HubbardToolBox} D. Jaksch and P. Zoller,
Annals of Physics {\bf 315}, 52  (2005).

   \bibitem{moss} S. Mossmann and C. Jung, Phys. Rev. A \textbf{74}, 033601 (2006).

   \bibitem{grae} E.\,M. Graefe, H.\,J. Korsch, and D. Witthaut, Phys. Rev. A \textbf{73}, 013617 (2006).
   
   \bibitem{buon} P. Buonsante, R. Franzosi, and V. Penna, Phys. Rev. Lett. \textbf{90}, 050404 (2003).
   
   \bibitem{stre} A.\,I. Streltsov, K. Sakmann, O.\,E. Alon, and L. \,S. Cederbaum, Phys. Rev. A \textbf{83}, 043604 (2011). 
   
   \bibitem{Stickney} J.\,A. Stickney, D.\,Z. Anderson, and A.\,A. Zozulya,
   Phys. Rev. A {\bf 75}, 013608 (2007). 
   
   \bibitem{laha} T. Lahaye, T. Pfau, and L. Santos, Phys. Rev. Lett. \textbf{104}, 170404 (2010).
   
   \bibitem{anna} L. Dell'Anna, G. Mazzarella, V. Penna, and L. Salasnich, Phys. Rev. A {\bf 87}, 053620 (2013). 
   
   \bibitem{chatterjee}  B. Chatterjee, I. Brouzos, L. Cao, and P. Schmelcher, 
   J. Phys. B: At. Mol. Opt. Phys. {\bf 46}, 085304 (2013).
   
   \bibitem{pete} D. Peter, K. Pawlowski, T. Pfau, and K. Rz{\c{a}}{\.{z}}ewski, J. Phys. B \textbf{45}, 225302 (2012).

   \bibitem{bara} M.\,A. Baranov, Phys. Rep. \textbf{464}, 71 (2008).
   
   \bibitem{laha1} T. Lahaye, C. Menotti, L. Santos, M. Lewenstein, and T. Pfau, Rep. Prog. Phys. \textbf{72}, 126401 (2009).
   



   \bibitem{adam} A. B{\" u}hler and H.\,P. B{\" u}chler, Phys. Rev. A \textbf{84}, 023607                   (2011).
   

 
   
   
   \bibitem{Bader} P. Bader and U.\,R. Fischer,
Phys. Rev. Lett. {\bf 103}, 060402 (2009).

   \bibitem{toma} T. Sowi{\'{n}}ski, O. Dutta, P. Hauke, L. Tagliacozzo, and M. Lewenstein, Phys. Rev. Lett. \textbf{108}, 115301 (2012).

   \bibitem{penr} O. Penrose and L. Onsager, Phys. Rev. \textbf{104}, 576 (1956).

   \bibitem{maik} A similar regular structure in a wavefunction sequence occurs also for dipolar bosons on an optical lattice ring in density-wave states, i.e., without first-order coherence, cf.\,\,M. Maik, P. Buonsante, A. Vezzani, and J. Zakrzewski, Phys. Rev. A \textbf{84}, 053615 (2011). 

   

\bibitem{gerbier} F. Gerbier {\it et al.}, 
Phys. Rev. Lett. {\bf 101}, 155303 (2008).
   
   
   \bibitem{hend} K. Henderson, C. Ryu, C. MacCormick and M.\,G. Boshier, New J. Phys. \textbf{11}, 043030 (2009).
   
   \bibitem{buecker} R. B\"ucker {\it et al.}, New J. Phys. {\bf 11}, 103039 (2009).
   
   \bibitem{mazz} G. Mazzarella, S.\,M. Giampaolo, and F. Illuminati, Phys. Rev. A \textbf{73}, 013625 (2006).

   \bibitem{voig} S. Kotochigova and E. Tiesinga, J. Chem. Phys. {\bf 123}, 174304 (2005).
   
   \bibitem{chi} C.-M. Chang, W.-C. Shen, C.-Y. Lai, P. Chen, and D.-W. Wang, Phys. Rev. A \textbf{79}, 053630 (2009).


\end{thebibliography}

\end{document}